\documentclass[aps,prc,reprint,amsmath,amssymb,showpacs,superscriptaddress]{revtex4-1}
\usepackage{graphicx}% Include figure files
\usepackage{dcolumn}% Align table columns on decimal point
\usepackage{bm}% bold math

\usepackage{hyperref}% add hypertext capabilities
\begin{document}
%\begin{CJK*}{UTF8}{} % Use default fonts from CJK (see below)

% Use the \preprint command to place your local institutional report
% number in the upper righthand corner of the title page in preprint mode.
% Multiple \preprint commands are allowed.
% Use the 'preprintnumbers' class option to override journal defaults
% to display numbers if necessary

%Title of paper
\title{Configuration interaction in symmetry-conserving covariant density functional theory}

\author{P. W. Zhao %({\CJKfamily{gbsn}} 赵鹏巍)
}
%\email{pwzhao@pku.edu.cn}
%\thanks{}
%\altaffiliation{}
\affiliation{Physics Division, Argonne National Laboratory, Argonne, Illinois 60439, USA}
\affiliation{State Key Laboratory of Nuclear Physics and Technology, School of Physics, Peking University, Beijing 100871, China}

\author{P. Ring 
}
\affiliation{Physik Department, Technische Universit\"at M\"unchen, D-85747 Garching, Germany}
\affiliation{State Key Laboratory of Nuclear Physics and Technology, School of Physics, Peking University, Beijing 100871, China}

\author{J. Meng 
}
\affiliation{State Key Laboratory of Nuclear Physics and Technology, School of Physics, Peking University, Beijing 100871, China}
\affiliation{School of Physics and Nuclear Energy Engineering, Beihang University, Beijing 100191, China}

%\date{\today}

\begin{abstract}
A new method to calculate spectroscopic properties of deformed nuclei is proposed: configuration interaction
on top of projected density functional theory (CI-PDFT). The general concept of this approach is discussed in the framework
of covariant density functional theory and its validity is illustrated in an application to the yrast band
of the nucleus $^{54}$Cr. It is found that the experimentally observed excitation energies for the yrast band in $^{54}$Cr
can be well reproduced. In contrast to conventional shell-model calculations, there is no core and only a relatively small number
of configurations is sufficient for a satisfying description. No new parameters are necessary, because the effective interaction
is derived from an universal density functional given in the literature.
\end{abstract}

% insert suggested PACS numbers in braces on next line
% 21.60.Jz Nuclear Density Functional Theory and extensions 
%21.10.Gv	Nucleon distributions and halo features
%21.30.-x	Nuclear forces
%21.10.-k Properties of nuclei; nuclear energy levels
%21.10.Re Collective levels
%21.60.Ev Collective models
%21.60.Cs	Shell model
%21.45.Ff	Three-nucleon forces
%23.20.-g Electromagnetic transitions
%23.20.Js Multipole matrix element
%27.20.+n  6 A 19
%27.40.+z 39  A  58
%27.60.+j 90  A 149
%27.50.+e 59  A  89
% insert suggested keywords - APS authors don't need to do this
%\keywords{}

%\maketitle must follow title, authors, abstract, \pacs, and \keywords
\maketitle

%\end{CJK*}

% body of paper here - Use proper section commands
% References should be done using the \cite, \ref, and \label commands

%\section{Introduction}
%=======================================================================================
Configuration interaction shell models~\cite{Caurier2005_RMP77-427} as well as
energy density functional theories (DFTs)~\cite{Bender2003_RMP75-121,Meng2016_IRNP10} have
been widely applied in the literature to the description of nuclear
properties. The shell model has the advantage of allowing the description of
all the spectroscopic properties of the system in the laboratory frame. The
physics contained in this approach is mainly expressed in terms of strong
mixing between different configurations, while a single configuration
contains little physics. The configuration mixing is determined by
effective interactions, which strongly depend on the choice of active shells
and various truncation schemes. Therefore, no universal shell-model
interaction can be used for all nuclei. Moreover, in order to preserve
rotational symmetry connected with the quantum numbers of angular momentum,
most of these calculations are carried out in a spherical basis. Within this
scheme, calculations for heavy deformed nuclei usually have to be carried out in large valence
spaces and, as a consequence, the dimensions of the corresponding matrices are
huge. This limits most of the applications of the shell-model approach for
heavy nuclei to spherical or nearly spherical systems.

Nuclear energy DFT has played an important role in the self-consistent
description of nuclei. In contrast to the configuration interaction shell
model, DFT is applicable for nuclei all over the nuclide chart as the complicated many-body problem
is mapped onto a single-particle problem, which is relatively easy to solve.
Over the years, several universal energy density functionals have been constructed, providing
a successful description with only a few phenomenological parameters.
They describe fruitful physics around the minima of the energy surfaces, and correlations are taken 
into account by breaking the essential symmetries. The disadvantage of this method is that, in the first
place, it is limited to nuclear ground states, and that many spectroscopic
properties are not accessible within this approach.

It is clear that the shell model and DFT are really two complementary
theoretical strategies to understand nuclear structure. Nuclear DFT is
applicable for the description of ground states and of bulk properties of
almost all nuclei in the nuclear chart, while shell-model calculations
reproduce fully quantum mechanically the spectroscopic details in specific
regions of the periodic table.
Therefore, for the goal of a unified and comprehensive description of all
nuclei, it is essential to develop a theory which combines the advantages from
both approaches. 

In the DFT regime, the generator coordinate method (GCM) has been
used to go beyond mean field and to study additional
correlations~\cite{Bender2003_RMP75-121,Niksic2011_PPNP66-519}. This approach has been implemented in
the nonrelativistic Skyrme~\cite{Valor2000Nucl.Phys.A145,Bender2006_PRC73-34322} and Gogny~\cite{Guzman2002Nucl.Phys.A201} functionals, and also the relativistic functionals~\cite{Niksic2006_PRC73-034308,Niksic2006_PRC74-064309}.  
Further developments considering the triaxial and the
octupole degrees of freedom are also available for both the nonrelativistic~\cite{Bender2008Phys.Rev.C024309,Rodriguez2010Phys.Rev.C064323,Rodriguez-Guzman2012Phys.Rev.C34336} and relativistic~\cite{Yao2015_PRC92-041304,ZHOU-EF2016_PLB753-227,Yao2009_PRC79-044312,Yao2010_PRC81-044311} models. 
Using the Gaussian overlap approximation (GOA), it can be 
considerably simplified through the derivation of a collective Hamiltonian, which 
is relatively simple for heavy  systems.
This method has been applied in Refs.~\cite{Libert1999_PRC60-054301,Yuldashbaeva1999_PLB461-1,Prochniak2004Nucl.Phys.A59} with nonrelativistic functionals and Refs.~\cite{Niksic2009_PRC79-034303,LI-Zhipan2009_PRC79-054301,LI-Zhipan2013_PLB726-866} with relativistic ones.
However, all these methods are limited to the description of very specific types of excitations (e.g., to quadrupole and/or octupole modes) and they can be applied only for relatively low energies, i.e., below the first two-quasiparticle excitations~\cite{Niksic2011_PPNP66-519}.

For high-spin spectra, the cranking model~\cite{Inglis1956_PR103-1786} has proven to be very practical and successful as it is a first-order approximation
for a variation after projection onto good angular momentum~\cite{Beck1970_ZP231-26}. 
Cranking models based on the DFTs have been widely used with the nonrelativistic functionals~\cite{Egido1993Phys.Rev.Lett.2876,Satula2005Rep.Prog.Phys.131} and the relativistic ones~\cite{Koepf1989_NPA493-61,Koepf1990_NPA511-279,Afanasjev1999_PRC60-051303}. 
Moreover, the tilted axis cranking relativistic
Hartree~\cite{Madokoro2000_PRC62-061301,PENG-J2008_PRC78-024313,ZHAO-PW2011_PLB699-181} and Hartree-Bogoliubov models~\cite{ZHAO-PW2015_PRC92-034319} have been developed and have achieved great success in many novel rotational
bands~\cite{Meng2013_FrontP8-55,ZHAO-PW2011_PRL107-122501,ZHAO-PW2012_PRC85-054310}. Similar approaches are also available with the nonrelativistic functionals~\cite{Olbratowski2004Phys.Rev.Lett.52501,Olbratowski2006Phys.Rev.C54308}.
However, the cranking models describe the physics only on average in a
rotating mean field. They mix configurations at constant angular velocity and
not at constant angular momentum. Therefore, level crossing phenomena are not described properly~\cite{Hamamoto1976_NPA271-15}. Attempts to include correlations due to symmetry breaking~\cite{Hara1982_NPA385-14} and to mix configuration beyond
the rotating mean field~\cite{Egido2016Phys.Rev.Lett.52502} are still in their infancy.

Therefore, we propose here a method which allows us to include quasiparticle
excitations, i.e., to mix collective and multiparticle excitations.
We call it configuration interaction-projected density functional theory (CI-PDFT).
In such a theory, we start from a successful density functional, and the self-consistent 
solution gives rise to a state in a minimum of
the energy surface. This state contains already important physics, and 
additional correlations, as required for spectroscopic studies, can be taken
into account with shell-model calculations in a configuration space built on
top of the minimum state from DFT. The dimension of such a configuration space
can be much smaller than that of a traditional shell-model calculation for
heavy nuclei, because the basis is not arbitrary, but optimized to the ground
state. The DFT solutions used here contain already important
correlations. Therefore, one can hope that such a method will be able in the
future to provide a global study of many nuclear properties with no additional
parameters beyond those of the well-established density functional.
Technically, this is reminiscent of the successful phenomenological
projected shell model (PSM) based on the Nilsson potential proposed by Hara
and Sun~\cite{Hara1995_IJMPE4-637,SUN-Y2009_PRC80-054306,SUN-Y2010_PRC82-031304}. 
This model, however, is limited to a few
valence shells. It is based on a pairing-plus-quadrupole
interaction and contains several phenomenological parameters,
which have to be adjusted in different mass regions.
In contrast, DFT is a more fundamental and
self-consistent theory based on more realistic two-body interactions.

%\section{Theoretical Framework}
%-----------------------------------------------------
In principle, this new method is applicable for all kind of density
functionals. Here, we concentrate on covariant density functional theory (CDFT), 
which has achieved great success in describing
many nuclear phenomena~\cite{Meng2016_IRNP10,Ring1996_PPNP37-193,Vretenar2005_PR409-101,Meng2006_PPNP57-470,Meng2013_FrontP8-55}.
The CDFT starts from a Lagrangian and the corresponding Kohn-Sham equations
derived from this Lagrangian have the form of a Dirac equation with effective
fields $S(\bm{r})$ and $V^{\mu}(\bm{r})$~\cite{Ring2016_IRNP10-1}. They are
driven in a self-consistent way by the densities and current distributions.
For open-shell nuclei, pairing correlations have to be taken into account. The
nuclear ground-state wave function $|\Phi\rangle$ corresponds to a
quasiparticle vacuum, where the quasiparticle basis can be defined by a
Bogoliubov transformation of a particle basis. Self-consistent solutions for
the single-quasiparticle Kohn-Sham equation yield the ground-state
wave function $|\Phi\rangle$; for details, see
Refs.~\cite{Niksic2011_PPNP66-519,Ring1996_PPNP37-193,Kucharek1991_ZPA339-23,ZHAO-PW2010_PRC82-054319}.

Starting from the ground state obtained in this way, one easily finds many
multiquasiparticle states, and they, together with the ground state, form the
configuration space. One
has to note that, here, rotational symmetry is violated for deformed nuclei; 
i.e., none of the states in the above configuration space is an eigenstate of
the angular momentum, and there are no good quantum numbers. This forbids a
satisfying description of spectroscopic properties and, thus, angular momentum
projection is required for these configurations~\cite{Ring1980}.

Suppose that we have chosen the set $\{|\Phi_{\kappa}\rangle\}$ of
multiquasiparticle states; the wave function $|\Psi\rangle$ in the laboratory
frame with good angular momentum quantum numbers $I$ and $M$ is obtained by
projection. For simplicity, we assume here axial symmetry of the basis; i.e.,
each of the multiquasiparticle components $|\Phi_{\kappa}\rangle$ is
characterized by a quantum number $K$, the projection of the angular momentum
onto the intrinsic symmetry axis. We thus find
\begin{equation}
|\Psi_{IM}^{\alpha}\rangle=\sum\limits_{\kappa}F_{\kappa}^{I\alpha}\hat
{P}_{MK}^{I}|\Phi_{\kappa}\rangle. \label{Eq.CollWav-PSM}
\end{equation}
Here, $\hat{P}_{MK}^{I}$ is the angular momentum projection
operator~\cite{Ring1980}, and the expansion coefficients $F_{\kappa}^{I\alpha}$ of
the eigenstate with the label $\alpha$ are determined by requiring that the
energy expectation value is stationary with respect to an arbitrary variation
of $F_{\kappa}^{I\alpha}$. This means that the Hamiltonian is diagonalized in
the shell-model subspace spanned by the nonorthogonal basis $\{\hat{P}%
_{MK}^{I}|\Phi_{\kappa}\rangle\}$, and this leads to the generalized
eigenvalue equations
\begin{equation}
\sum\limits_{\kappa^{\prime}}\{H_{\kappa\kappa^{\prime}}^{I}-E_{\alpha}%
^{I}N_{\kappa\kappa^{\prime}}^{I}\}F_{\kappa^{\prime}}^{I\alpha}=0,
\label{Eq.PSM-axia}
\end{equation}
with the Hamiltonian overlap matrix $H_{\kappa\kappa^{\prime}}^{I}$ and the
norm overlap matrix $N_{\kappa\kappa^{\prime}}^{I}$:%
\begin{equation}
H_{\kappa\kappa^{\prime}}^{I}=\langle{\Phi_{\kappa}}|{\hat{H}\hat
{P}_{KK^{\prime}}^{I}}|{\Phi_{\kappa^{\prime}}}\rangle,\quad N_{\kappa
\kappa^{\prime}}^{I}=\langle{\Phi_{\kappa}}|{\hat{P}_{KK^{\prime}}^{I}}%
|{\Phi_{\kappa^{\prime}}}\rangle. \label{Eq.HN}%
\end{equation}
The weight functions $F_{\kappa}^{I\alpha}$ are normalized by the condition%
\begin{equation}
\sum\limits_{\kappa\kappa^{\prime}}F_{\kappa}^{I\alpha}N_{\kappa\kappa
^{\prime}}^{I}F_{\kappa^{\prime}}^{I\alpha^{\prime}}=\delta_{\alpha\alpha^{\prime}}.
\end{equation}
Here, $\hat{H}$ represents the residual interaction~\cite{Ring1980}, which, in
the present framework of relativistic density functional theory, is found as
the second derivative of the energy density functional with respect to the
density matrix~\cite{Daoutidis2009_PRC80-024309}:
\begin{equation}
V_{nmn^{\prime}m^{\prime}}^{{}}=\frac{\delta^{2}E[\hat{\rho}]}{\delta\hat
{\rho}_{nm^{{}}}\delta\hat{\rho}_{n^{\prime}m^{\prime}}}.
\label{Eq.interaction}
\end{equation}
Since the exchange terms are neglected in the mean-field calculations, for
self-consistency, we should also neglect exchange terms in the present beyond
mean-field level. Moreover, as the first step, we also neglect the
contribution from the spacelike components of the vector channel in the
present work due to the limitation of numerical costs.

As usual for generalized eigenvalue problems of the form (\ref{Eq.CollWav-PSM}
), the basis functions $\hat{P}_{MK}^{I}|\Phi_{\kappa}\rangle$ are not
orthogonal and, therefore, the coefficients $F_{\kappa}^{I\alpha}$ cannot be
interpreted as probability amplitudes for the state $\kappa$. As discussed in
Ref.~\cite{Ring1980}, these probability amplitudes $G_{\kappa}^{I\alpha}$ are
written as
\begin{equation}
G_{\kappa}^{I\alpha}=\sum\limits_{\kappa\kappa^{\prime}}\left(  N^{I}\right)
_{\kappa\kappa^{\prime}}^{1/2}F_{\kappa^{\prime}}^{I\alpha}. \label{GI_kappa}
\end{equation}

Once the weights $F_{\kappa}^{I\alpha}$ of the nuclear collective wave
functions $|\Psi_{IM}^{a}\rangle$ in Eq.~(\ref{Eq.CollWav-PSM}) are known, it
is straightforward to calculate all physical observables, such as the
electromagnetic transition probabilities. The reduced transition probability
for a transition between an initial state $I_{i}$ and a final state 
$I_{f}$ is defined by%
\begin{equation}
B(E2;I_{i},\alpha\rightarrow I_{f},\alpha^{\prime})=\frac{e^{2}}{2I_{i}%
+1}\left\vert \left\langle \Psi_{I_{f}}^{\alpha^{\prime}}||\hat{Q}_{2}%
||\Psi_{I_{i}}^{\alpha}\right\rangle \right\vert ^{2}.
\end{equation}
Note that here $e$ denotes the bare value of the proton charge, and there is
no need to introduce effective charges.

%\section{Results and discussion}
%-----------------------------------------------------
In this paper, the results of the newly developed CI-PDFT are illustrated by
taking the nucleus $^{54}$Cr as an example. In order to simplify the
evaluation of the Hamiltonian and the norm overlap, we used the relativistic
point coupling Lagrangian PC-PK1~\cite{ZHAO-PW2010_PRC82-054319}. This avoids
 the complicated overlap integrals for functionals with finite range.
As the first application, the particle number projection is not included. 
Therefore, we employ an approximate scheme for correcting the mean value of the
particle number as in Ref~\cite{Yao2011Phys.Rev.C014308}. It has also been
found in numerous applications that the simple phenomenological PSM~\cite{Hara1995_IJMPE4-637} 
is very successful without particle number projection. 
The single-nucleon wave functions are solved via an expansion in a basis of
cylindrical harmonic oscillator wave functions with $N_F=12$ major shells,
and the basis parameters are $\hbar\omega_{0}=41A^{-1/3}$ and $\beta
_{0}=0$~\cite{Gambhir1990_APNY198-132,Niksic2014_CPC185-1808}. Pairing
correlations are treated by the BCS method with the density-independent
$\delta$ force given in Ref.~\cite{ZHAO-PW2010_PRC82-054319}. 
We have checked that the oscillator expansion has converged well for $N_F=12$.
It is found that the lowest energy state has a prolate
deformation of around $\beta_{2}=0.24$. This is in agreement with
the experimental value $\beta_2=0.247 \pm 0.067$ of Ref.~\cite{Pritychenko2016At.DataNucl.DataTables1}. 

The evaluation of the Hamiltonian and the norm overlap integrals
in Eq.~(\ref{Eq.HN}) can be carried out in particle space (see, for instance,
Ref.~\cite{Niksic2006_PRC73-034308}) or in quasiparticle space. In this work, we used the latter method following the expressions of Ref.~\cite{Hara1995_IJMPE4-637} for separable effective interactions.
For this purpose, we expressed the interaction terms in the zero-range functional PC-PK1 as a sum of separable terms in a similar way as it has been done in Refs.~\cite{Daoutidis2009_PRC80-024309,Daoutidis2011Phys.Rev.C44303} for the zero-range functionals
PC-F1~\cite{Burvenich2002_PRC65-044308} and DD-PC1~\cite{Niksic2008_PRC78-034318}.

\begin{figure}[tbh]
\centering
\includegraphics[width=8cm]{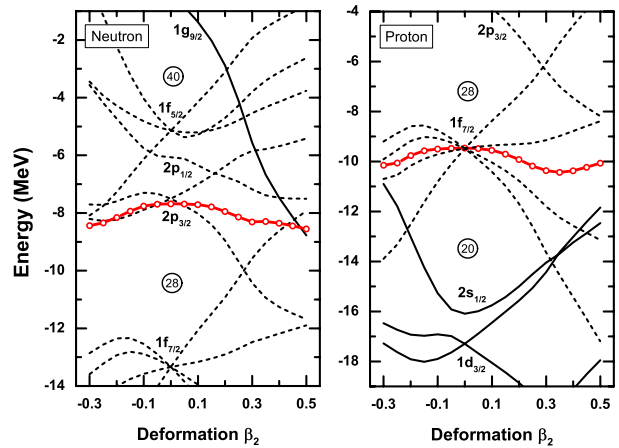}
\caption{(Color online) The neutron (left
panel) and proton (right panel) single-particle levels for $^{54}$Cr, as
functions of the quadrupole deformation $\beta_{2}$. The dashed and solid
lines represent the levels with negative and positive parities, respectively.
The thick lines with open circles denote the position of the Fermi energy.}%
\label{fig1}
\end{figure}

The neutron and proton single-particle levels are presented in Fig.~\ref{fig1} as
functions of the quadrupole deformation $\beta_{2}$. It is important to
identify the single-particle states near the Fermi surfaces, which take part
in the low-lying quasiparticle excitations. In the present work, the
configuration space consists of the zero- and two-quasiparticle states
\begin{equation}
|0\rangle,\quad\alpha_{\nu^{{}}}^{\dagger}\alpha_{\nu^{\prime}}^{\dagger
}|0\rangle,\quad\alpha_{\pi^{{}}}^{\dagger}\alpha_{\pi^{\prime}}^{\dagger
}|0\rangle,
\end{equation}
and the dimension of this configuration space is truncated with an cutoff in
the two-quasiparticle excitation energy $E_{\mathrm{cut}}=6.5$ MeV for both
protons and neutrons. The corresponding configuration space consists of 37
states including 18 two-quasineutron, 18 two-quasiproton excited states, and
the quasiparticle vacuum $|0\rangle$. The convergence of the dimension of the
configuration space has been tested by comparing with the results obtained
with a larger energy cutoff $E_{\mathrm{cut}}=7.0$ MeV, and it turns out that
such a configuration space is sufficient to obtain convergent results for the
states with $I\leq10$ in $^{54}$Cr.

\begin{figure}[tbh]
\centering
\includegraphics[width=7cm]{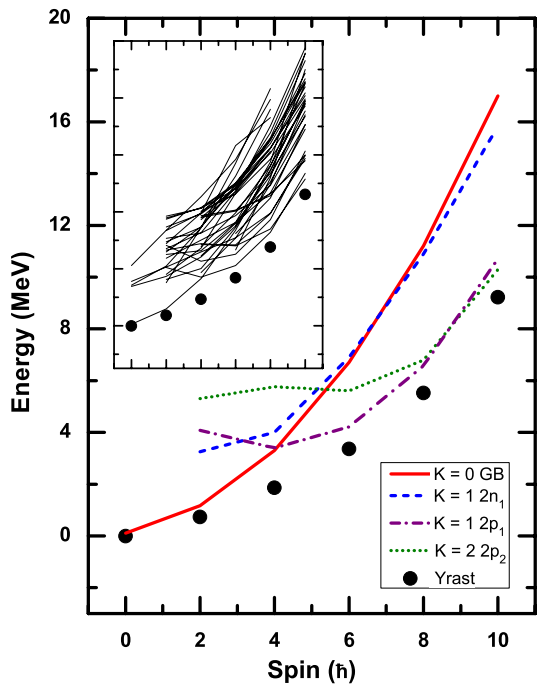}\caption{(Color online) Theoretical yrast
band (solid circles) and the angular momentum projected states for some
important configurations which are explained in Table~\ref{table1} in $^{54}%
$Cr. Inset: Angular momentum projected states for all the configurations
considered together with the yrast band (solid circles). Note that the energy
of the yrast state $0^{+}$ is renormalized to zero. }%
\label{fig2}%
\end{figure}

In the inset of Fig.~\ref{fig2}, the energies of angular momentum projected
states for all the 37 configurations contained in the present calculation are
shown as functions of the angular momentum together with the yrast band. Such
a figure is also called a band diagram~\cite{Hara1995_IJMPE4-637}, in which
the rotational behavior of each configuration, as well as its relative energy
compared to other configurations, is clearly visualized. Therefore, one has to
trace the configuration of each band in the band diagram. Since axial
symmetry is imposed in the present work, the $K$ value (the projection of
angular momentum on the symmetry axis of the deformed system) can be easily
used to identify the configurations. Note that the angular momentum for each
configuration does not always start from zero due to the fact that the
starting angular momentum $I$ should satisfy the relation $I\geq|K|$. Moreover,
the starting angular momentum for each configuration is also not necessarily
equal to the spin of the corresponding bandhead, since in some cases, as shown
in Fig.~\ref{fig2}, the band energies are not always increasing with the spin.

For even-even nuclei, the zero quasiparticle ground band has $K=0$, whereas a
two quasiparticle band has a $K$ given by the coupling of the $K$ values of
its constituent quasiparticles. A superposition of them imposed by
configuration mixing gives the final results at each spin. The ensemble of the lowest calculated states with the energy $E_{0}^{I}$ at each spin forms the yrast band, which is shown
by solid circles in Fig.~\ref{fig2}.

\begin{table*}[htb]
\caption{Important configurations with their quasiparticle excitation energies
and the amplitudes $G^{I}_{\kappa}$ in the yrast state. Here, we list only the
configurations whose contributions to the yrast state are larger than 1\%.}%
\label{table1}%
\centering
\begin{tabular*}
{1.0\textwidth}[c]{@{\extracolsep{\fill} }cccccccccc}\hline
& $E$ & $K$ & Configurations & 0 & 2 & 4 & 6 & 8 & 10\\\hline
gs & 0.00 & 0 & -- & 0.959 & 0.856 & 0.623 & 0.280 & 0.150 & 0.113\\
2$n$1 & 2.68 & 1 & $(2p_{3/2})_{k=1/2}\otimes(1f_{5/2})_{k=1/2}$ &  & 0.314 &
0.448 & 0.241 & 0.098 & 0.100\\
& 3.36 & 1 & $(2p_{3/2})_{k=1/2}\otimes(2p_{3/2})_{k=-3/2}$ &  & 0.225 &
0.308 & 0.164 & 0.055 & 0.000\\
& 4.64 & 2 & $(2p_{3/2})_{k=1/2}\otimes(1f_{5/2})_{k=3/2}$ &  & -0.044 &
-0.146 & -0.076 & -0.037 & -0.064\\
& 4.64 & 1 & $(2p_{3/2})_{k=1/2}\otimes(1f_{5/2})_{k=-3/2}$ &  & 0.068 &
0.126 & 0.085 & 0.037 & 0.028\\\hline
& 2.39 & 0 & $(1f_{7/2})_{k=5/2}\otimes(1f_{7/2})_{k=-5/2}$ & 0.265 & 0.146 &
-0.084 & -0.232 & -0.228 & -0.166\\
2$p$1 & 2.55 & 1 & $(1f_{7/2})_{k=3/2}\otimes(1f_{7/2})_{k=-5/2}$ &  & 0.224 &
0.430 & 0.521 & 0.400 & 0.341\\
& 2.55 & 4 & $(1f_{7/2})_{k=3/2}\otimes(1f_{7/2})_{k=5/2}$ &  &  & 0.013 &
0.205 & 0.183 & 0.146\\
& 2.71 & 0 & $(1f_{7/2})_{k=3/2}\otimes(1f_{7/2})_{k=-3/2}$ & -0.055 &
-0.028 & 0.020 & 0.283 & 0.297 & 0.280\\
2$p$2 & 3.56 & 2 & $(1f_{7/2})_{k=1/2}\otimes(1f_{7/2})_{k=-5/2}$ &  & -0.047 &
-0.127 & -0.386 & -0.416 & -0.409\\
& 3.56 & 3 & $(1f_{7/2})_{k=1/2}\otimes(1f_{7/2})_{k=5/2}$ &  &  & -0.018 &
-0.270 & -0.320 & -0.332\\
& 3.71 & 1 & $(1f_{7/2})_{k=1/2}\otimes(1f_{7/2})_{k=-3/2}$ &  & 0.076 &
0.159 & -0.088 & -0.277 & -0.256\\
& 3.71 & 2 & $(1f_{7/2})_{k=1/2}\otimes(1f_{7/2})_{k=3/2}$ &  & -0.043 &
-0.075 & 0.070 & 0.178 & 0.152\\
& 4.42 & 1 & $(1f_{7/2})_{k=5/2}\otimes(1f_{7/2})_{k=-7/2}$ &  & -0.152 &
-0.142 & -0.019 & 0.020 & 0.061\\
& 4.42 & 6 & $(1f_{7/2})_{k=5/2}\otimes(1f_{7/2})_{k=7/2}$ &  &  &  & -0.130 &
-0.069 & -0.054\\
& 4.57 & 2 & $(1f_{7/2})_{k=3/2}\otimes(1f_{7/2})_{k=-7/2}$ &  & 0.009 &
-0.073 & -0.180 & -0.204 & -0.227\\
& 4.57 & 5 & $(1f_{7/2})_{k=3/2}\otimes(1f_{7/2})_{k=7/2}$ &  &  &  & 0.194 &
0.216 & 0.192\\
& 5.58 & 3 & $(1f_{7/2})_{k=1/2}\otimes(1f_{7/2})_{k=-7/2}$ &  &  & 0.032 &
0.148 & 0.286 & 0.367\\
& 5.58 & 4 & $(1f_{7/2})_{k=1/2}\otimes(1f_{7/2})_{k=7/2}$ &  &  & -0.002 &
-0.152 & -0.251 & -0.355\\\hline
\end{tabular*}
\end{table*}

It is usually not necessary to put too many bands in a single band diagram, but
to select instead some of the most important ones. In Fig.~\ref{fig2}, the theoretical
band diagram for some important configurations are given together with
the yrast band; these different bands can be distinguished by different line
styles. It has been found that, as the nucleus starts rotating, the energy of
the ground-state band increases rapidly toward the higher-energy region. However,
the two quasiparticle bands rise slowly with spin and, therefore, they can
cross the ground-state band. Here, we use $2n_{1}$, $2p_{1}$, and $2p_{2}$ to
label the relevant configurations, and the detailed information is listed in
Table~\ref{table1}.

\begin{figure}[htb]
\centering
\includegraphics[width=7cm]{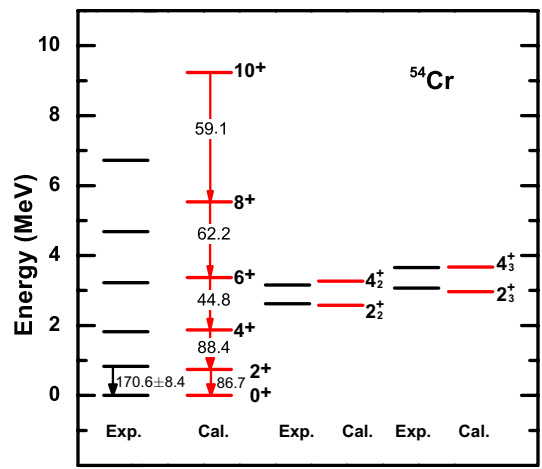}\caption{(Color online) Comparison between the
calculated energy levels (MeV) and $B(E2;\downarrow)$ transition probabilities ($e^2~\rm{fm}^4$) and the corresponding data available in Refs.~\cite{Pritychenko2016At.DataNucl.DataTables1,Devlin1999_PRC61-017301} for the
yrast band with the angular momentum $I\le10$, and also the nonyrast $I=2,\,4$ states.}%
\label{fig3}%
\end{figure}

After configuration mixing by diagonalizing the residual interaction in the
projected configuration space, the yrast states are obtained (solid circles in
Fig.~\ref{fig2}), and these are the theoretical results to be compared with the
data. In Fig.~\ref{fig3}, the calculated energy levels and the
$B(E2;\downarrow)$ transition probabilities are given in comparison with the available
data~\cite{Devlin1999_PRC61-017301} for the yrast band and also the nonyrast $I=2,\,4$ states.

In Table~\ref{table1}, we show, for each of the yrast states with angular
momentum $I$, the corresponding decomposition into the basis configurations. These
configurations are characterized by their quasiparticle energies
$E=E_{1}+E_{2}$, their $K$ values, and their approximate quantum numbers in the
spherical basis. The remaining columns represent the amplitudes $G_{\kappa
}^{I}$ of the yrast states in Eq.~(\ref{GI_kappa}). As expected, it is found
that the ground state with $I=0$ has only $K=0$ admixtures. It corresponds
essentially to the mean-field ground state $|0\rangle$. With increasing
angular momentum, we observe admixtures of higher $K$ values, in particular of
the $K=1$ two-neutron state $2n_{1}$, the $K=1$ two-proton state $2p_{1}$, and
the $K=2$ two-proton state $2p_{2}$. This is in agreement with the
band diagram in Fig.~\ref{fig2}.

It is found that the calculated energies for yrast $I=2$,
$4$, $6\hbar$ states, and also the nonyrast $I=2$, $4$ states are in good agreement with the data. However, the calculated $B(E2;2^+\rightarrow 0)$ transition probability
is roughly a factor two smaller than the experimental value. This might have to do with the
fact that, as seen in Table~\ref{table1}, besides the quasiparticle vacuum $|0\rangle$, admixtures of 2-neutron configurations mostly contribute to the 2$^+$ state wave function, and these do not contribute to the $B(E2;2^+\rightarrow 0)$ value.
For the higher spin states with $I=8$, $10\hbar$, the calculated band is too extended in comparison with
the data, or in other words, the calculated moments of inertia are too small.
The reason for this type of deviation could be the neglect of the time-odd
components and, thus, the omission of the mixed currents in the present
calculations. In fact, it has been shown that the time-odd components would
influence the higher spin states in the previous investigation of rotational
bands with cranking CDFT (see, e.g.,
Refs.~\cite{Koepf1990_NPA511-279,LIU-Lang2012_SCG55-2420}). 
In this context, it was found that the inclusion of time-odd components can
increase the moments of inertia and, thus, lead to more compressed rotational bands.
On the other hand, it is possible that the
influence of four-quasiparticle configurations should be taken into account. In fact, investigations within the phenomenological
projected shell model in this region of the periodic table~\cite{SUN-Y2010_PRC82-031304}, which include
also four-quasiparticle configurations, exhibit better agreement with the experimental data, but
at the cost of phenomenological parameters. In order to answer such questions, 
more detailed investigations within CI-PDFT are definitely
necessary.

%\section{Summary}
%-----------------------------------------------------
We proposed a new method, called CI-PDFT, for the description of spectra of
deformed nuclei all over the periodic table. Starting from a successful
density functional, a minimum in the energy surface is found and a number of
quasiparticle configurations is determined. Together with the deformed ground
state, they are projected to restore the symmetries broken at the mean field
level, and these projected states form the space for configuration mixing
calculations using the effective interaction derived
from the underlying energy density functional. In this way, one keeps all the
advantages of the mean field theory, and goes beyond mean field through
mixing of states with good symmetries, which is a concept of the standard
shell model. This method will provide a tool for a global study of many
nuclear properties with no parameters beyond those of the well-established
underlying density functional.

A computer code was developed for relativistic point coupling models and first
calculations have been carried out for the Lagrangian PC-PK1. Pairing
correlations are taken into account with a $\delta$ force. The method has been
tested for the $^{54}$Cr nucleus. It has been
found that the ground state has a quadrupole deformation $\beta_{2}=0.24$. 
By using this minimum as a basis for the projection and after configuration mixing by diagonalizing the
residual interaction in a sufficient projected configuration space, the
experimental yrast band in $^{54}$Cr can be reproduced in the low spin part of 
the band, while the calculated band is too extended for the high-spin
part. The present investigation is quite simple. It can be improved in several
ways: The effects of time-odd components in the residual interaction should be
considered carefully and higher configurations with more quasiparticles could
be included. The present model also cannot describe collective vibrations. In
principle, this is possible by considering a larger number of
two-quasiparticle states in the basis. Work in this direction is in progress.

We note that our model may also be used to describe the mixing of different shapes
and shape coexistence phenomena, if the shapes are just obtained by a reoccupation of
two particles, as it happens in many light nuclei.  A more detailed investigation along this 
direction is required in the future. Moreover, if one considers only the band heads 
of collective bands, e.g., $\gamma$ bands, the quasiparticle random 
phase approximation (QRPA) will be certainly a much simpler model, and it would 
be interesting to compare QPRA with our method. However, the QRPA approaches 
based on a fixed minimum will not be able to describe collective high-spin states, 
for which one would need QRPA based on cranked mean-field states (see, e.g., Ref.~\cite{Egido1980Nucl.Phys.A1}).

% If you have acknowledgments, this puts in the proper section head.
\begin{acknowledgments}
We thank R. V. F. Janssens and Robert B. Wiringa for careful reading of the manuscript.
This work is supported by U.S. Department of Energy (DOE), Office of Science, Office of Nuclear Physics, under Contract No. DE-AC02-06CH11357, by the Chinese Major State 973 Program 2013CB834400, and by the NSFC Grant No. 11335002 and No. 11621131001.
We also acknowledge support from the Laboratory Computing Resource Center at Argonne National Laboratory, and the DFG
Cluster of Excellence ``Origin and Structure of the
Universe'' (www.universe-cluster.de).
\end{acknowledgments}

%\bibliography{paper}

%merlin.mbs apsrev4-1.bst 2010-07-25 4.21a (PWD, AO, DPC) hacked
%Control: key (0)
%Control: author (8) initials jnrlst
%Control: editor formatted (1) identically to author
%Control: production of article title (-1) disabled
%Control: page (0) single
%Control: year (1) truncated
%Control: production of eprint (0) enabled
%

\end{document}